\documentclass[twocolumn,showpacs,preprintnumbers,amsmath,amssymb,pre,unsortedaddress,superscriptaddress]{revtex4}
\usepackage{graphicx}% Include figure files
\usepackage{dcolumn}% Align table columns on decimal point
\usepackage{bm}% bold math

\begin{document}

%\preprint{APS/123-QED}

\title{Evolving network models under a dynamic growth rule}
\author{Ke Deng}
\affiliation{Department of Physics, Jishou University, Jishou
 Hunan, 416000, China}
\affiliation{Department of Physics, Xiangtan University, Xiangtan
Hunan 411105, China}

\author{Ke Hu}
\affiliation{Department of Physics, Xiangtan University, Xiangtan
Hunan 411105, China}

\author{Yi Tang}
\affiliation{Department of Physics, Xiangtan University, Xiangtan
Hunan 411105, China}

\date{\today}

\begin{abstract}
Evolving network models under a dynamic growth rule which
comprises the addition and deletion of nodes are investigated. By
adding a node with a probability $P_a$ or deleting a node with the
probability $P_d=1-P_a$ at each time step, where $P_a$ and $P_d$
are determined by the Logistic population equation, topological
properties of networks are studied. All the fat-tailed degree
distributions observed in real systems are obtained, giving the
evidence that the mechanism of addition and deletion can lead to
the diversity of degree distribution of real systems. Moreover, it
is found that the networks exhibit nonstationary degree
distributions, changing from the power-law to the exponential one
or from the exponential to the Gaussian one. These results can be
expected to shed some light on the formation and evolution of real
complex real-world networks.
\end{abstract}

 \maketitle

\section{INTRODUCTION}
Considerable interest is focused on complex networks currently due
to their potential to describe many systems in nature and society
\cite{1,2}. Theoretical models are developed to reproduce
topological properties of these systems \cite{3,4,5}. The recently
observed \emph{scale-free} (SF) property \cite{5} has made the
scientists to realize that \emph{static} network models \cite{3,4}
do not provide appropriate descriptions to real systems which
essentially keep \emph{growing} with time \cite{1,2}. Barab\'{a}si
and Albert \cite{5} proposed a simple evolving network model (BA
model) to explain such SF property. In this model, the growing
nature of real systems is captured by a BA-type \emph{growth
rule}. According to this rule, one node is added into the network
at each time step, intending to mimic the growing process of real
systems. Another ingredient in this model is the mechanism of
\emph{preferential attachment} (PA), which assumes that newly
added nodes are attached preferentially to nodes with higher
degrees \cite{6}. Based on the BA-type growth rule, many evolving
network models are introduced \cite{1,2}. The modeling framework
of network evolution is formed.

Among the many quantities proposed to characterize topological
properties of networks\cite{1,2}, the \emph{degree distribution}
$p(k)$, which gives the probability that a node in the network
possesses $k$ edges, is of particular importance. The study of
networks has undergone a transition from the investigation of
static network models with Poisson degree distribution \cite{3,4}
to the explanation of real systems with various \emph{fat-tailed
degree distributions} (FTDDs) \cite{1,2}. It was found that many
real-world networks are characterized by the \emph{power-law
degree distribution} (PLDD) \cite{5}, as well as the
\emph{exponential degree distribution} (EDD) \cite{19}. Some more
exhaustive experiment reveals that, in addition to the PLDD and
the EDD, the \emph{truncated power-law degree distribution}
(TPLDD) \cite{9,20} and the \emph{truncated exponential degree
distribution} (TEDD) \cite{21} are also observed. Furthermore, in
Ref. \cite{9}, the \emph{Gaussian degree distribution} (GDD) was
reported as well.

One theoretical challenge has been to explain the origin of these
observed FTDDs. In the BA model, the fact is revealed that growing
networks with PA yield the PLDD \cite{5}. In addition, growing
networks without PA were also studied in the random evolving
network model (REN model) \cite{7,22}, in which networks grow by
the growth rule of BA-type, while the newly added nodes connect to
\emph{randomly chosen} existing ones. In this model, the EDD was
obtained. In order to explain those observed degree distributions
which are neither strict power-law nor strict exponential
\cite{9,20,21}, many other mechanisms were added into this two
models, such as the mechanism of adding and rewiring edges between
existing nodes \cite{8,23,ad7,ad8}, the mechanism of ageing and
cost \cite{9}, as well as the mechanism of information filtering
\cite{20}. These more specialized models indeed created the TPLDD
or the EDD in some parameter regimes. However, relatively lesser
attention is paid to systematical studies on the origin of the
diversity of the observed FTDDs, with some recent exceptions
\cite{9,a,ad10}.

The BA-type growth rule gives a somewhat simplified description to
the evolution of real system. As a matter of fact, in real growing
networks, there are constant addition of new elements, but
accompanied by permanent removal of old elements (deletion of
nodes) \cite{12,13,ad12}. The scaling behavior of growing networks
has already been shown to be strongly affected when the deletion
of node is taken into account \cite{d,e,ad9,ad11}. In a recent
work \cite{a}, we proposed a new type of network growth rule which
comprises the addition and deletion (AD) of nodes. Based on such
AD growth rule, by adding a node with a probability $P_a$ or
deleting a node with probability $P_d=1-P_a$ at each time step,
topological properties of growing networks with and without PA are
studied, respectively. In ref. \cite{a}, we considered a simple
case: $P_a$ and $P_d$ were treated as constants. ($P_a$ is an
adjustable parameter in the model.) All the observed FTDDs of real
systems were obtained in the model, indicating that the mechanism
of AD can lead to the diversity of FTDD in real systems.

The AD growth rule introduced in \cite{a} gives a more detailed
description to the evolution of real systems. However, this
evolution can be even more complex. For example, as real systems
grow, limited resources will cause interelement competition which,
in turn, has a tendency to retard the growth of these systems.
Such competition is intensified when the number of element is
increased. As a result, growth rate of real systems usually
decreases with the increase of the system size, exhibiting the
so-called \emph{density-dependent growth} \cite{16}. In fact,
competitive dynamics has been found to dominate the evolution of
variety of social \cite{26}, biological \cite{27} and economic
\cite{28} network systems. Thus, with respect to the AD growth
rule, to be more realistic, $P_a$ should be a decreasing function
of the number of nodes in the network, rather than a constant
\cite{a}. In this paper, we introduce the \emph{Logistic
population equation} \cite{16} into the AD growth rule. As a
result, the probabilities of addition (deletion) becomes a
decreasing (increasing) function of the network size. Based on
this dynamical AD growth rule, growing networks models with and
without PA are investigated, respectively. In the present models,
networks exhibit the density-dependent growth and all the observed
FTDDs are created. Moreover, it is found that the networks exhibit
\emph{nonstationary} degree distributions, changing from power-law
to exponential one or from exponential to Gaussian one. These
results can be expected to shed some light on the formation and
evolution of real complex real-world networks.

\section{NETWORK MODELS WITH LOGISTIC AD GROWTH RULE}
The Logistic equation \cite{16} was proposed by Pierre Verhulst
for the analysis of population competitive dynamics. Verhulst
assumed that during the population growth, as a result of the
competition for the limited resources, the death rate per
individual is a lineally increasing function of the number of
population. Then the Logistic equation can be written as
\begin{equation}
\label{eq.1} \
\frac{dN(t)}{dt}=[a-cN(t)]N(t)=a[1-\frac{N(t)}{K}]N(t) \;,
\end{equation}
where $N(t)$ denotes the number of population at time $t$; $a$ and
$c$ are constants; $a$ is the birth rate per individual and
$cN(t)$ stands for the death rate per individual at time $t$;
$K=a/c$ is called \emph{carrying capacity} \cite{17,18} which
represents the largest population the environment can support due
to limited resources.

A more realistic description to the real-network's evolution can
be achieved by the introduction of such Logistic dynamics into the
AD growth rule. To do this, we obtain the probability of addition
$P_a$ and the probability of deletion $P_d$ by the following two
equations
\begin{equation}
\label{eq.2} \ \frac{P_a}{P_d}=\frac{a N(t)}{c N(t)^2}
\end{equation}
and
\begin{equation}
\label{eq.3}
 P_a+P_d=1,
\end{equation}
which yield
\begin{equation}
\label{eq.4} P_a(t)=\frac a{a+c N(t)}=\frac K{K+N(t)}
\end{equation}
and
\begin{equation}
\label{eq.5} P_d(t)=\frac{c N(t)}{a+c N(t)}=\frac {N(t)}{K+N(t)}.
\end{equation}
Where $N(t)$ stands for the number of nodes at time $t$ and $K$ is
the parameter which denotes the maximum nodes in the network.

Then the Logistic AD model can be defined as follows: We start
from $m_0$ isolated nodes, which act as the nuclear of a growing
network. At each time step, either a new node is added to the
network with probability $P_a$ or a randomly selected old node is
removed from the network with probability $P_d=1-P_a$, where $P_a$
and $P_d$ are determined by Eq.~(\ref{eq.4}) and Eq.~(\ref{eq.5}),
respectively.

When a new node is added to the network, there are still two ways
for it to attach to the existing nodes in the network. One way is
to randomly choose $m$ nodes to set up connections (growing
network without PA) \cite{7,22}, and the other way is to
preferentially select $m$ nodes to connect by means of
preferential probability in the BA model \cite{5}, which reads
\begin{equation}
\label{eq.6} \ \Pi _i=\frac{k_i+1}{\sum_j\left( k_j+1\right) },
\end{equation}
where $k_i$ is the degree of the $i$th node (growing network with
PA). Here, we should note that in order to give chance for
isolated nodes to receive a new edge we choose $\Pi _i$
proportional to $k_i+1$ \cite{8}.

One can find from Eq.~(\ref{eq.4}) and Eq.~(\ref{eq.5}) that when
$N(t)\ll K$, $P_a(t)\gg P_d(t)$. This means that the networks grow
rapidly at the initial stages of their evolution. In fact, in a
Logistic model \cite{16}:
\begin{equation}
\label{eq.7} \frac{dN(t)}{dt}=a \left[ \frac{K-N(t)}K\right] N(t),
\end{equation}
when $N(t)\ll K$,
\begin{equation}
\label{eq.8} \frac{dN(t)}{dt}\approx a N(t),
\end{equation}
the solution is:
\begin{equation}
\label{eq.9} N\left( t\right) =C_{0}\exp \left( a t\right),
\end{equation}
where $C_{0}$ is a integral constant. Indeed, this kind of
exponential growth has been observed in many newly emerged
real-world networks, such as the World-Wide-Web (WWW) and the
Internet \cite{12,13}. This indicates that the rapid growth of
some real systems in their young age is well described by the
Logistic AD model. As $N(t)$ increase, $P_a(t)$ decrease while
$P_d(t)$ increase, the growth of network is slowed down. In the
limit of large $t$,
\begin{equation}
\label{eq.10} \lim\limits_{t\rightarrow \infty
}P_a(t)=\lim\limits_{t\rightarrow \infty }P_d(t)=\frac 12,
\end{equation}
and
\begin{equation}
\label{eq.11} \lim\limits_{t\rightarrow \infty }N\left( t\right)
=K,
\end{equation}
i.e., the network reach a steady state and the number of nodes has
the upper limit $K$. The above analysis imply that networks in the
Logistic AD model exhibit density-dependent growth characterizing
the evolution of many real systems, not captured in most previous
network models.

\begin{figure}
\includegraphics[scale=0.6]{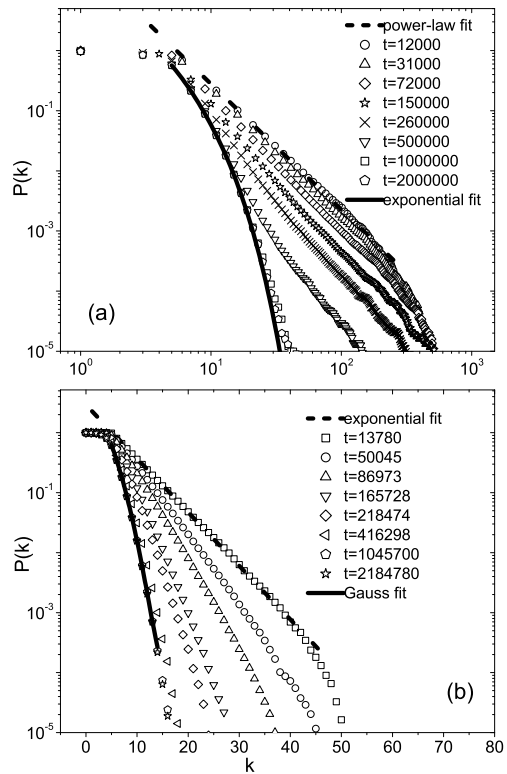}
\caption{\label{lwith/out} (a) Cumulative degree distribution of
the growing network with PA, for different time steps $t$, in
logarithmic scales. The dash line is power-law fit and the solid
line is exponential fit. (b) Cumulative degree distribution of the
growing network without PA, for different time steps $t$, in
semi-logarithmic scales. The dash line is exponential fit and the
solid line is Gaussian fit. Results in both (a) and (b) are based
on many independent realizations.}
\end{figure}

We investigated the degree distribution of the networks by
extensive computer simulations. In the simulation, we set
$m_0=m=5$ and $K=100000$. Network is left to evolve until the
steady state is reached. Cumulative degree distributions of
growing networks with and without PA at different time step are
given in Fig.~\ref{lwith/out}(a) and Fig.~\ref{lwith/out}(b),
respectively. We found that in the Logistic AD model, the networks
exhibit nonstationary degree distribution before the steady state
is reached. This is illustrated in Fig.~\ref{lwith/out}. For the
growing network with PA, as time goes on, $P(k)$ of the network
undergos a process of transition, which can be roughly separated
into several stages: $(1)$ At the earlier stage of network
evolution, i.e., when $t\leq 72000$, the network exhibits various
PLDDs with different power-law exponents. In addition, the
exponent increases with time. Particularly, in the asymptotic
cases of $P_{a}$ being close to 1, the network is almost
equivalent to the well-known BA model, thus it have PLDD with
power-law exponents $\gamma=3$. $(2)$ $72000<t<1000000$. $P(k)$ of
the network is truncated by an exponential cutoff and the network
exhibits the TPLDD during this stage. Generally, if a node needs
to obtain the high degree, it must exists in network with enough
long time. While, longer do the nodes live, higher is the
probability that they are deleted. Moreover, with $t$ increasing,
$P_{d}$ increases. Thus the surviving probability of the nodes
with high degrees is greatly reduced, and the degree distribution
evolves with time, gradually changing from a power-law form to a
power-law with a exponential cutoff, then to a exponential one.
$(3)$ In the limit of large $t$, e.g., when $t\geq 1000000$, the
network shows a well shaped stationary EDD [see
Fig.~\ref{lwith/out}(a)]. Speciously, this asymptotic case seems
to be a non-growing network (or a very slowly growing one). The
case of the preferential attachment on a non-growing network was
considered in Ref. \cite{12} where it was found that $P(k)$ is not
stationary, changing from a power-law type directly to is a
Gaussian one. On the other hand, for the growing network without
PA, as time goes on, $P(k)$ of the network exhibits a continuous
transition from the EDD to a variety of TEDDs which prove to be a
series of intermediate states, and in the limit of large $t$,
e.g., when $t\geq 1045700$, the network exhibits a well shaped
stationary GDD [see Fig.~\ref{lwith/out}(b)].

The results of Fig.~\ref{lwith/out} reveal a nonstationary
behavior for the degree distribution of real systems. During the
evolution of these systems, various events take place on different
timescales. These events include, for instance, the addition and
deletion of nodes, the creating and rewiring of edges between
existing nodes (internal edges), and so on. In general, the
timescale on which a node join or leave the network (the addition
and deletion of nodes) may be much longer than the timescale on
which other events take place. For example, in a social network,
creating internal edges means that the individuals make new
friends, which happens on the timescale that can be as short as
hours or days. While the timescale on which individuals are born
or die is typically some years or decades. So that our model,
which is based on the mechanism of addition and deletion, gives
the prediction of a long-run behavior of real systems. For this
reason, this kind of nonstationary FTDD of real systems, which can
be difficult to observe in a relatively short time interval, has
been neglected in most previous studies \cite{1,12}. However, in
recent years, the Internet and the WWW is in the initial and rapid
growth stage of their evolutions. Perhaps, the observations for
their degree distributions might provide some clues for our
results. The WWW at the document level had grown at least five
times larger during the two years time delay between the first and
last web crawl, and the degree distributions obtained at different
time are reported to be of a power-law form \cite{ad1,ad2,ad3}.
The power-law exponent, though seems to be invariable for the
in-degree distributions, has an increasing tendency with the
sample size or time for the out-degree distribution, changing from
2.45 in 1999 to 2.72 in 2000 \cite{ad1,ad2,ad3}. Another clue is
refereed to the Internet on the inter-domain level. The degree
distribution of the Internet is also of power-law form. The
exponent, however, suffer a little change from 2.15 in November
1997, to 2.16 in April 1998, then to 2.20 in December 1998
\cite{ad4}. In fact, it is hard to achieve this precision. One may
estimate the value of the exponent using the highest degrees and
the system's size \cite{ad5}. Such estimations also confirm the
reported values. For November 1977, we gets $\gamma\simeq 2.22$,
for April 1998, $\gamma\simeq 2.24$, and for December 1998,
$\gamma\simeq 2.26$. Of course, exist additional factors
(nonlinear PA \cite{ad5} and accelerating growth of edges
\cite{ad6}, etc.) that may change these values. Finally, we are
glad to point out that in a recent work \cite{23}, such
nonstationary degree distribution was also reported when the
creation of internal edges was considered in growing and static
networks.

It is well known that the structure of complex networks has strong
effects on their function. Therefor such neglect, at least in some
case, may be dangerous because networks with different degree
distribution act differently if we consider the dynamic processes
taking place on them. This is especially the case, for example, in
the strategy designing for the prevention of virus spreading or
for the defense of network attacking \cite{b,c,11}. Provided that,
after some years, the WWW has gradually changed into an
exponential network, but is still treated as a scale-free one,
what will happen to this network of high technological importance
if it runs under the antivirus strategy designed for a scale-free
network?

\section{CONCLUSION}
In summary, we have proposed a dynamic growth rule which comprises
the addition and deletion of nodes in the network. Based on this
rule, by adding a node with a probability $P_a$ or deleting a node
with probability $P_d=1-P_a$ at each time step, topological
properties of growing networks with and without PA are studied
respectively. The probabilities of addition and deletion are
assumed to be determined by the Logistic population equation. In
our model, networks exhibit the density-dependent growth which
characterizes the evolution of various real systems, and all the
observed degree distributions of real systems are created.
Moreover, the networks exhibit nonstationary degree distributions,
changing from power-law to exponential one or from exponential to
Gaussian one. It is discussed that this kind of nonstationary
behavior can be unconspicuous in real-world networks. The result
indicates that the degree distribution of real systems, which has
been believed to be stationary in most previous studies, may in
fact be nonstationary.

\section*{ACKNOWLEDGMENTS}
This work was supported by the Outstanding Young Researcher's
Foundation of Hunan Province, China, Grant No. 03JJY1001, and by
the Natural Science Foundation of Hunan Province, China, Grant No.
00JJY2072. This work is also supported by the Foundation of
Educational Committee of Hunan Province, china, Grant No. 00C189
and No.01B019.

\end{document}